\documentclass[letters,fleqn,usenatbib]{mnras}

\usepackage{newtxtext,newtxmath}
\usepackage[T1]{fontenc}
\usepackage{ae,aecompl}

\usepackage{graphicx}
\usepackage[usenames, dvipsnames]{color}
\usepackage{subfigure}

\usepackage[load-configurations=abbreviations]{siunitx}
\DeclareSIUnit\year{yr}
\DeclareSIUnit\au{au}
\DeclareSIUnit\parsec{pc}
\DeclareSIUnit\solarmass{\ensuremath{\textrm{M}_\odot}}
\DeclareSIUnit\solarradius{\ensuremath{\textrm{R}_\odot}}

\newcommand{\gc}{globular cluster}

\newcommand{\ngc}[1]{NGC~#1}

\newcommand{\target}{target star}
\newcommand{\Target}{Target star}
\newcommand{\Po}{\SI{1}{\day}}
\newcommand{\Pf}{\SI{51}{\day}}
\newcommand{\Pe}{\SI{83}{\day}}
\newcommand{\Ph}{\SI{167}{\day}}
\newcommand{\BHmass}{\SI{4.36(41)}{\solarmass}}

\newcommand{\Autoref}[1]{%
 \begingroup%
 \renewcommand\figureautorefname{Figure}%
 \renewcommand\equationautorefname{Equation}%
 \autoref{#1}%
 \endgroup%
}

\def\figureautorefname{Fig.}

\title[Black hole candidate in \ngc{3201}]{A detached stellar-mass black hole candidate in the \gc{} \ngc{3201}}

\author[B. Giesers et al.]{%
Benjamin Giesers,$^{1}$\thanks{E-mail: giesers@astro.physik.uni-goettingen.de}
Stefan Dreizler,$^{1}$\thanks{E-mail: dreizler@astro.physik.uni-goettingen.de}
Tim-Oliver Husser,$^{1}$
Sebastian Kamann,$^{1,\,2}$\newauthor
Guillem Anglada Escud\'e,$^{3}$
Jarle Brinchmann,$^{4,\,5}$
C. Marcella Carollo,$^{6}$\newauthor
Martin M. Roth,$^{7}$
Peter M. Weilbacher,$^{7}$
Lutz Wisotzki$^{7}$
\\
$^{1}$Institut f\"ur Astrophysik, Georg-August-Universit\"at G\"ottingen, Friedrich-Hund-Platz 1, 37077 G\"ottingen, Germany\\
$^{2}$Astrophysics Research Institute, Liverpool John Moores University, 146 Brownlow Hill, Liverpool L3 5RF, United Kingdom\\
$^{3}$School of Physics and Astronomy, Queen Mary University of London, 327 Mile End Road, London, United Kingdom\\
$^{4}$Leiden Observatory, Leiden University, P.O. Box 9513, 2300 RA, Leiden, The Netherlands\\
$^{5}$Instituto de Astrof{\'\i}sica e Ci{\^e}ncias do Espa\c{c}o, Universidade do Porto, CAUP, Rua das Estrelas, PT4150-762 Porto, Portugal\\
$^{6}$Institute for Astronomy, Swiss Federal Institute of Technology (ETH Zurich), CH-8093 Zurich, Switzerland\\
$^{7}$Leibniz-Institut f\"ur Astrophysik Potsdam (AIP), An der Sternwarte 16, 14482 Potsdam, Germany\\
}

\date{Accepted XXX. Received YYY; in original form ZZZ}

\pubyear{2017}

\begin{document}
\label{firstpage}
\pagerange{\pageref{firstpage}--\pageref{lastpage}}
\maketitle

\begin{abstract}
As part of our massive spectroscopic survey of 25 Galactic globular
clusters with MUSE, we performed multiple epoch observations of \ngc{3201}
with the aim of constraining the binary fraction. In this cluster, we found one
curious star at the main-sequence turn-off with radial velocity variations
of the order of \SI{100}{\kilo\meter\per\second}, indicating the membership
to a binary system with an unseen
component since no other variations appear in the spectra. Using an
adapted variant of the generalized Lomb-Scargle periodogram, we could
calculate the orbital parameters and found the companion to be a detached
stellar-mass black hole with a minimum mass of \BHmass{}. The result is
an important constraint for binary and black hole evolution models in
globular clusters as well as in the context of gravitational wave sources.
\end{abstract}


%%%%%%%%%%%%%%%%%%%%%%%%%%%%%%%%%%%%%%%%%%%%%%%%%%

\section{Introduction}
Owing to their old ages and high masses, Galactic globular clusters probably have produced
a large number of stellar-mass black holes during their lifetimes. 
Nevertheless, there is an ongoing debate about the number of black holes that actually remain in the cluster.
In the absence of continuous star formation, stellar-mass black holes soon become the most massive objects in the cluster, where they accumulate around the centres as a consequence of mass segregation. However, because of the high mass-ratio with respect to the surviving low-mass stars ($\gtrsim4:1$), it is expected that the black holes form a dense nucleus that is decoupled from the dynamics of the remaining cluster \citep{spitzer1969}. Interactions within this nucleus are then expected to eject the majority of black holes, so that only few survive after \SI{1}{\giga\year} \citep{kulkarni1993,sigurdsson1993}.

However, over the past years, radio observations have revealed several sources in extragalactic and Galactic \gc{}s that are likely to be stellar-mass black holes according to their combined radio and X-ray properties \citep{maccarone2007,strader2012,chomiuk2013}. Under the assumption that only a small fraction of the existing black holes are actively accreting matter from a companion \citep{kalogera2004}, these detections point to much richer black hole populations in \gc{}s than previously thought. In fact, state-of-the art models for clusters do predict that the retention fractions of black holes might be significantly enhanced compared to the earlier studies mentioned above \citep[e.g.][]{breen2013,morscher2013}. The reason for this is that the \citet{spitzer1969} instability only develops partially and the black hole nucleus does not detach from the kinematics of the remaining cluster. As a consequence, the evaporation time-scale is prolonged.

The search for black holes in \gc{}s has recently gained further importance through the first detection of gravitational waves, produced by the coalescence of two massive black holes \citep{abbott2016a}. As shown by \citet{abbott2016b} or \citet{askar2017}, dense star clusters represent a preferred environment for the merging of such black hole binaries. Hence, it would be crucial to overcome the current observational limits in order to increase our sample of known black holes. Compared to radio or X-ray studies, dynamical searches for stellar companions have the advantages of also being sensitive to non-accreting black holes and of providing direct mass constraints. We are currently conducting a large survey of 25 Galactic \gc{}s with MUSE \citep[Multi Unit Spectroscopic Explorer,][]{muse}, which provides us with spectra of currently \num{600} to \num{27000} stars per cluster \citep[see][]{kamann2017}. Our survey includes a monitoring for radial velocity variations, which is very sensitive to the detection of stellar companions of massive objects (i.e. black holes, neutron stars and white dwarfs). Here, we report the detection of a \BHmass{} black hole in the \gc{} \ngc{3201}.

\section{Observations and data reduction}
\label{sec:observations}

\begin{table}
 \caption{Barycentric corrected radial velocity $v_\textrm{r}$, (Vega) magnitude $I_\textrm{F814W}$, and seeing (See.) measurements for the \target{}. ($\Delta$JD: Julian observation date $\textrm{JD} - \SI{2456978}{\day}$, ESO ID: ESO Programme ID Code.)}
 \label{table:observations}
 \centering
 \setlength\tabcolsep{3pt}
 \begin{tabular}{r c c c c}
 \hline\hline
     $\Delta$JD [d] &  $v_\textrm{r}$ [km/s] & $I_\textrm{F814W}$ [mag] & See. [\arcsec] & ESO prog. ID \\
  \hline
  0.83733 & \num{570.8(22)} & \num{16.88(8)} & 0.82 & 094.D-0142 \\
  11.85794 & \num{512.9(27)} & \num{16.89(6)} & 0.90 & 094.D-0142 \\
  11.86438 & \num{511.9(28)} & \num{16.91(8)} & 0.84 & 094.D-0142 \\
  30.81091 & \num{475.0(22)} & \num{16.83(6)} & 0.60 & 094.D-0142 \\
  30.83250 & \num{482.1(24)} & \num{16.88(7)} & 0.66 & 094.D-0142 \\
  31.78096 & \num{477.1(20)} & \num{16.86(6)} & 0.72 & 094.D-0142 \\
  31.80265 & \num{480.8(21)} & \num{16.87(7)} & 0.88 & 094.D-0142 \\
  149.49256 & \num{536.8(20)} & \num{16.90(6)} & 0.60 & 095.D-0629 \\
  151.47767 & \num{550.8(36)} & \num{16.82(7)} & 1.12 & 095.D-0629 \\
  153.47698 & \num{559.3(25)} & \num{16.86(8)} & 1.00 & 095.D-0629 \\
  156.47781 & \num{585.4(22)} & \num{16.86(8)} & 1.04 & 095.D-0629 \\
  160.47808 & \num{609.5(20)} & \num{16.87(8)} & 0.70 & 095.D-0629 \\
  441.74475 & \num{476.2(20)} & \num{16.90(6)} & 0.66 & 096.D-0175 \\
  441.76738 & \num{472.0(18)} & \num{16.88(7)} & 0.64 & 096.D-0175 \\
  443.74398 & \num{474.8(21)} & \num{16.89(6)} & 0.66 & 096.D-0175 \\
  443.76792 & \num{472.5(21)} & \num{16.85(8)} & 0.54 & 096.D-0175 \\
  538.47410 & \num{471.1(19)} & \num{16.86(6)} & 0.80 & 097.D-0295 \\
  542.47958 & \num{471.9(20)} & \num{16.88(6)} & 0.78 & 097.D-0295 \\
  808.87270 & \num{501.4(27)} & \num{16.80(6)} & 0.60 & 098.D-0148 \\
  809.87675 & \num{512.7(25)} & \num{16.87(8)} & 0.96 & 098.D-0148 \\
  \hline
 \end{tabular}
\end{table}

The observational challenge in \gc{}s is the crowding resulting in
severe blending of nearby stars especially in the cluster cores. For photometric
measurements of dense \gc{}s, instruments like the Hubble Space
Telescope (HST) provide sufficient spatial resolution to obtain
independent measurements of most of the stars \citep{acs1}. For spectroscopic
surveys, investigations of dense stellar fields were limited to the
brightest stars or to regions with sufficiently reduced crowding.
In \gc{}s, the combination of field of view (\SI{1}{\arcmin}x\SI{1}{\arcmin})
and spatial sampling (300x300 spaxel$^2$)
of MUSE allows us to extract spectra with a spectral
resolution of $1800 < R < 3500$ of some thousand stars per
exposure \citep[for more details see][]{husser2016}.
We use the standard ESO MUSE pipeline to reduce the MUSE raw
data \citep{weilbacher2012}. The extraction is done with a PSF-fitting technique using
the combined spatial and spectral information \citep{kamann,kamann2014}. 
% At this step, we also subtract the unresolved stellar background as well as the night sky emission. 

% The overall aim of our stellar census in \gc{}s with MUSE
% is twofold.  On the one hand we will investigate the spectroscopic
% properties of more than half a million of cluster members down to the
% main-sequence. On the other hand we will investigate the dynamical
%properties of \gc{}s.
This work is part of our investigation of the
binarity of clusters using the radial velocity method. Binaries that
are able to survive in the dense environment of a \gc{} are so tight \citep[see][]{hut1992}
that even HST is unable to resolve the single components such that they
will appear as a single point-like source. Radial velocity surveys,
however, will rather easily detect those compact binary
systems. Except for the case when both stars have the same brightness,
the extracted MUSE spectrum will be dominated by one of the stars,
i.e. we mostly detect single line binary systems. With the knowledge of
the stellar mass and the orbital parameters from our analyses, we can
nevertheless infer the minimum masses of the unseen companions.

Up to this publication, we have observed three \gc{}s \ngc{104}
(47 Tuc), \ngc{3201}, and \ngc{5139} ($\omega$ Cen) with sufficiently
many observations to analyse the radial velocity
signals $v_\textrm{r}$ of individual stars. We identified three
stars with radial velocity variations exceeding \SI{100}{\kilo\meter\per\second}.
Only one of them (hereafter called \target{}) in \ngc{3201} appears in two
overlapping pointings, resulting in 20 extracted spectra with good signal-to-noise ratios
(from \num{24} to \num{56}, \num{43} on average). 
The other two stars need more observations to analyse the orbital parameters.
\Autoref{table:observations} lists the \target{}'s derived radial velocities and
seeing values for each observation from five observation runs.

\section{Photometric and spectroscopic analysis}
\label{sec:validation}

The investigation of the \target{} properties is performed in two steps.
Photometry provides the mass of the \target{}
and allows us to exclude alternative explanations for the large radial velocity signal.
%and allows to exclude alternative explanations 
%such as pulsations for the large
%radial velocity signal other than caused by a binary with a massive unseen companion.
Below, we describe the spectral analysis from which we obtain the cluster membership. 

\subsection{Photometric analysis}
\label{sec:photometric}
%\begin{figure}
%\centering
%\resizebox{0.95\hsize}{!}{\includegraphics{plots/bh_finding_chart_wide}}
%\caption{Finding chart with the \target{} marked within a red
%  circle. The cluster centre is marked by a red X \citep[position
%  from][]{harris}. The image is taken from the HST ACS globular
%  cluster survey from \citet{acs1, acs2}. The distance from the
%  \target{} to the cluster centre is \SI{10.8}{\arcsec}.} 
%\label{fig:finding_chart}
%\end{figure}
\begin{figure}
\centering
\subfigure[HST]{\includegraphics[width=0.47\hsize]{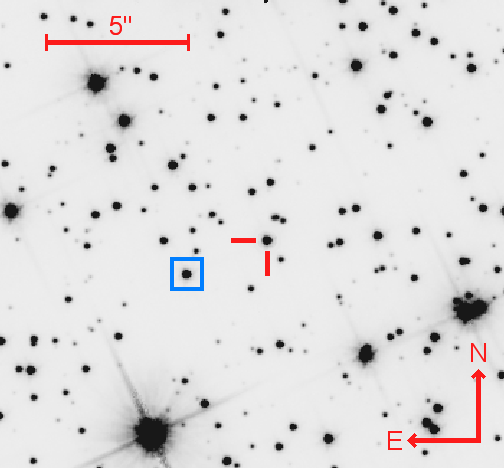}}\quad
\subfigure[MUSE]{\includegraphics[width=0.485\hsize]{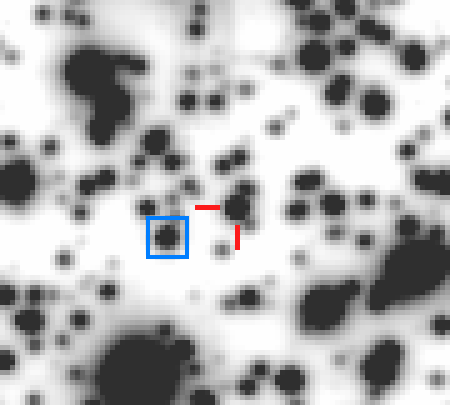}}
\caption{Charts with the \target{} marked with red crosshairs and the reference star marked within a blue square.
  (a) The image is taken from the HST ACS globular cluster survey from \citet{acs1} and \citet{acs2}.
  (b) Same field of view seen by MUSE with a seeing of \SI{0.6}{\arcsec}.
  The displayed image is a cut (90x81 pixels) from an integrated white light image of the MUSE data cube.}
\label{fig:finding_chart}
\end{figure}

\Autoref{fig:finding_chart} shows the \target{} (red crosshairs) with a
magnitude of $I_\textrm{F814W} = \num{16.87}$ mag (Vega) and its nearby stars from
HST ACS data \citep{acs2}. To monitor the reliability of our analyses,
we have picked the star in the blue square with a magnitude of
$I_\textrm{F814W} = \num{17.11}$ as a reference star ($\textrm{RA} = 10^\mathrm{h}\;17^\mathrm{m}\;37\fs36$, $\textrm{Dec.} = -46\degr\;24\arcmin\;56\farcs48$).
The three nearest stars have magnitudes fainter than \num{19} mag.
For a consistency check and for investigating photometric variability, we
convolve each flux calibrated spectrum with the corresponding ACS filter function and
compare it to the ACS photometry. For all extracted spectra, we reach a
magnitude accuracy of at least 95 per cent compared to the HST
ACS magnitude (see \autoref{table:observations}).
This indicates that the \target{} is extracted without contamination.

\begin{figure}
\resizebox{\hsize}{!}{\includegraphics[clip, trim=0.3cm 0.4cm 0.4cm 0.4cm]{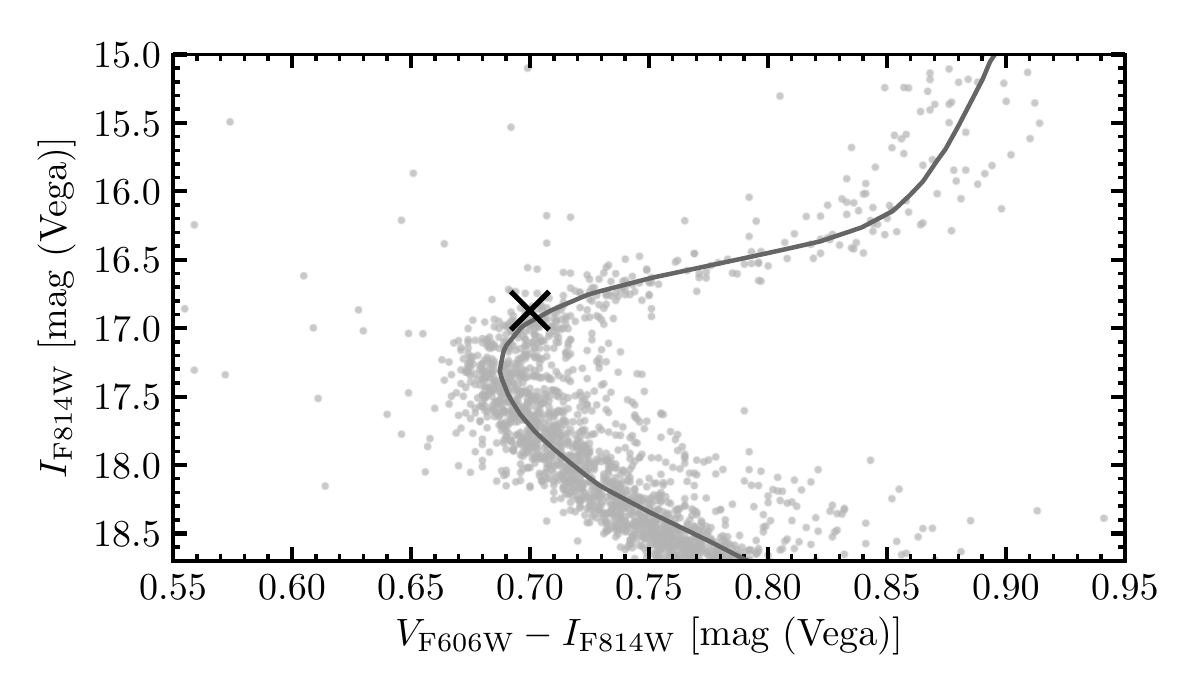}}
\caption{CMD of the \gc{} \ngc{3201} created
  from the HST ACS photometry of \citet{acs2}. The \target{} position is highlighted by a black X.
  The line represents our best-fitting PARSEC isochrone \citep[for more details see \autoref{sec:photometric}]{parsec}.}
\label{fig:ngc3201_CMD_with_observed_star}
\end{figure}

In order to derive the mass and surface gravity $\log{g}$ for the
spectrum fit (see \autoref{table:target_star}) of the \target{}, we
compare its HST ACS colour and magnitude
with a PARSEC isochrone \citep{parsec}. For the \gc{}
\ngc{3201}, we found the best matching isochrone compared to the
whole HST ACS Colour-magnitude diagram (CMD) with the isochrone parameters
[M/H]~=~\SI{-1.39}{dex} \citep[slightly above the comparable literature value $\lbrack\mathrm{Fe/H}\rbrack$ = \SI{-1.59}{dex}, ][]{harris},
age~=~\SI{11}{\giga\year}, extinction $E_{B-V}$~=~\num{0.26}, and
distance~=~\SI{4.8}{\kilo\parsec} (see
\autoref{fig:ngc3201_CMD_with_observed_star}). The \target{} is at the 
main-sequence turn-off with a mass of \SI{0.81(5)}{\solarmass}
as estimated from the isochrone.

% \begin{figure}
% \resizebox{\hsize}{!}{\includegraphics{plots/1406607_params_photometry}}
% \caption{Differential photometry for all observations obtained with
%   MUSE for the \target{}. 20 reference stars with an intrinsic standard
%   deviation of $0.005$ mag (systematic uncertainty) were used.}
% \label{fig:1406607_params_photometry}
% \end{figure}

Although the position of the target star in the CMD is not in the
classical instability strip, we want to exclude radial velocity variations
caused by photometric variability (e.g. due to pulsations). Therefore,
we use the differential photometry method on our MUSE data.
We first run an iterative algorithm to find a large
sample of reference stars (which are present in all
observations). After these stars have been identified, normally with
an intrinsic standard deviation of $0.005$ mag in the filter
$I_\textrm{Bessel}$ reconstructed from the extracted spectrum,
we compare each star of each observation with our
reference stars selecting 20 with comparable colour. 
%To get the
%uncertainties to the resulting magnitude differences we use
%\SI{68}{\percent}-percentiles on signal to noise bins as estimated
%standard deviations.
The photometric analysis for the \target{} shows
no significant variation (see \autoref{table:observations}). We would
expect significant changes in the brightness of the \target{} to
explain such a high radial velocity amplitude via pulsations. We
therefore conclude that the radial velocity variations are not
intrinsic to the \target{}. The absence of large photometric 
variations also suggests the absence of an interacting binary. 
We like to note that no radio or X-ray source is known at the \target{}'s position.
For instance, \citet{strader2013} performed a deep systematic
radio continuum survey for black holes also in \ngc{3201},
but did not publish any discovery. Some X-ray sources were found
by \citet{webb2006}, but none at the \target{}'s coordinates.

\subsection{Spectroscopic analysis}
After the extraction of point sources, the individual spectra of the 20
visits are fitted against our \textit{G\"ottingen Spectral Library} \citep{husser}
of synthetic PHOENIX spectra to determine the stellar effective temperature
and metallicity as well as radial velocities and the telluric absorption
spectrum. The simultaneous fit of stellar and telluric spectra is
performed with a least-squares minimization comparing the complete
observed spectrum against our synthetic spectra. The telluric absorptions are
used to correct for small remaining wavelength calibration
uncertainties, allowing us to reach a radial velocity precision down to
\SI{1}{\kilo\meter\per\second}. For more details about the stellar parameter fitting methods,
we refer to \citet{husser,husser2016}. We find huge changes in
radial velocity between the spectra of up to \SI{138}{\kilo\meter\per\second}.
We got the same variation if we do just cross-correlations between the spectra
and the initial template, of course with higher uncertainties.
Besides the radial velocity signal, the spectral fitting of the
individual spectra did not reveal any other significant variations.
The radial velocities are given in \autoref{table:observations}, whereas the mean stellar
parameters are given in \autoref{table:target_star}.

\begin{figure}
\resizebox{\hsize}{!}{\includegraphics[clip, trim=0.3cm 0.4cm 0.4cm 0.4cm]{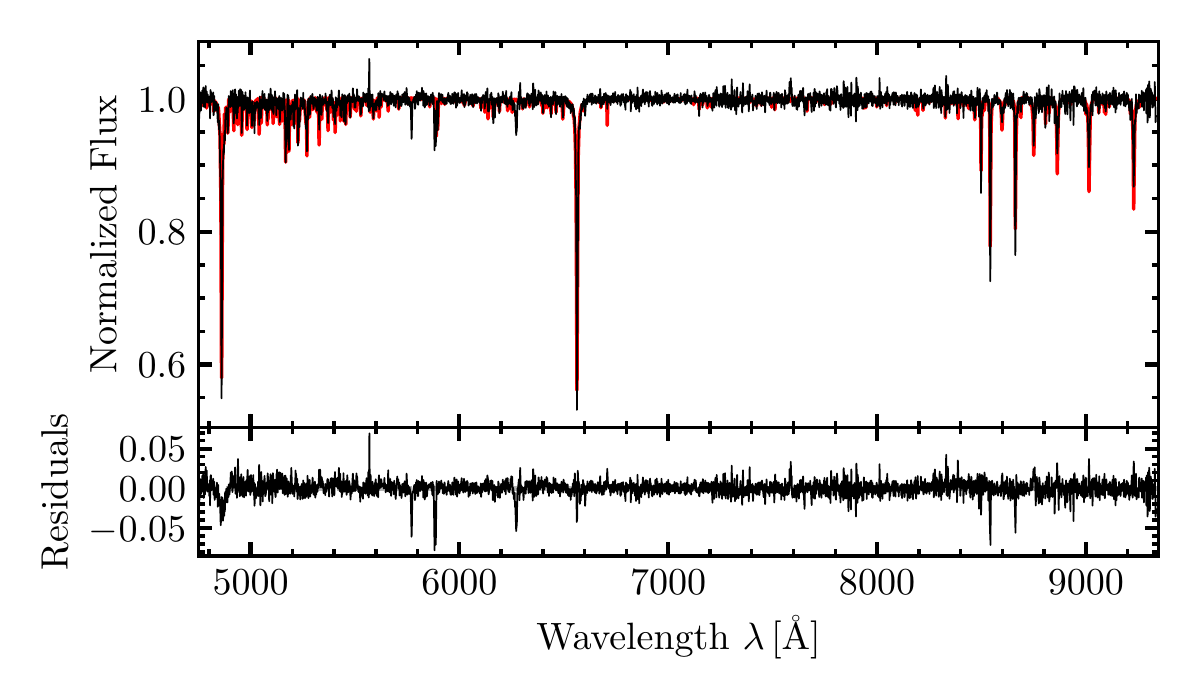}}
\caption{The combination of all radial velocity corrected spectra of the \target{} (in black). The best-fitting PHOENIX spectrum is indicated in the background in red (thicker curve for better visibility). The bottom panel shows the
residuals after subtracting the best fit to the data.
%The resulting $S/N$ is about \num{160} per \SI{1}{\angstrom}.
}
\label{fig:spectrum} 
\end{figure} 

Using the derived radial velocities, the individual spectra are placed
into rest frame and are then combined using the drizzling method from
\citet{fruchter2002} and finally normalized to the continuum.  The
combined spectrum in \autoref{fig:spectrum} is compared to one of our
PHOENIX spectra for a star at the main-sequence turn-off in the \gc{}
\ngc{3201}. The spectral properties match the position of the star in the
CMD displayed in \autoref{fig:ngc3201_CMD_with_observed_star}. We note that there
are no emission lines that could indicate, for example, a cataclysmic variable or a
compact binary with an illuminated low-mass star and a hot companion
like a white dwarf or neutron star.

\begin{table}
 \caption{\Target{} properties. The position and magnitude were taken
   from the HST ACS globular cluster survey catalogue \citep{acs2},
   mass and surface gravity are derived from isochrone fitting,
   effective temperature and metallicity are derived from spectral
   fitting. For more details see \autoref{sec:validation}.}
 \label{table:target_star}
 \centering
 \setlength\tabcolsep{3pt}
 \begin{tabular}{r l r l}
 \hline\hline
 RA & $10^\mathrm{h}\;17^\mathrm{m}\;37\fs090$ & Dec. & $-46\degr\;24\arcmin\;55\farcs31$ \\
 $I_\textrm{F814W}$ & \num{16.87(2)} mag (Vega) & \\
 $M$ & \SI{0.81(5)}{\solarmass} & $\log{g}$ & \SI{3.99(5)}{dex} \\
 $T_{\textrm{eff}}$ & \SI{6126(20)}{\kelvin} & \textrm{[M/H]} & \SI{-1.50(2)}{dex} \\
 \hline
 \end{tabular}
\end{table}

%\begin{table}
 %\caption{\Gc{} \ngc{3201} properties taken from the \citet{harris} catalogue.}
 %\label{table:ngc3201}
 %\centering
 %\begin{tabular}{r l}
 %\hline\hline
 %RA & 10\degr 17\arcmin 36\farcs82 \\
 %Decl & -46\degr 24\arcmin 44\farcs9 \\
 %Distance to Sun & \SI{4.9}{\kilo\parsec} \\
 %Barycentric radial velocity & \SI{494.0(2)}{\kilo\meter\per\second} \\
 %Central velocity dispersion & \SI{5.0(2)}{\kilo\meter\per\second} \\
 %Metallicity [Fe/H] & \SI{-1.59}{dex} \\
 %\hline
 %\end{tabular}
%\end{table}

The mean radial velocity \SI{506(1)}{\kilo\meter\per\second} and mean metallicity
$\textrm{[M/H]} = \SI{-1.50(2)}{dex}$ of the \target{} is in good agreement with
the cluster parameters in the \citet{harris} catalogue. Further, the fitted
radial velocity of the binary barycentre (see \autoref{table:binary_results})
matches precisely the radial velocity of the cluster \SI{494.0(2)}{\kilo\meter\per\second}.
This makes the \target{} a bonafide cluster member.

\section{Results}
\label{sec:results}
\begin{figure}
\resizebox{\hsize}{!}{\includegraphics{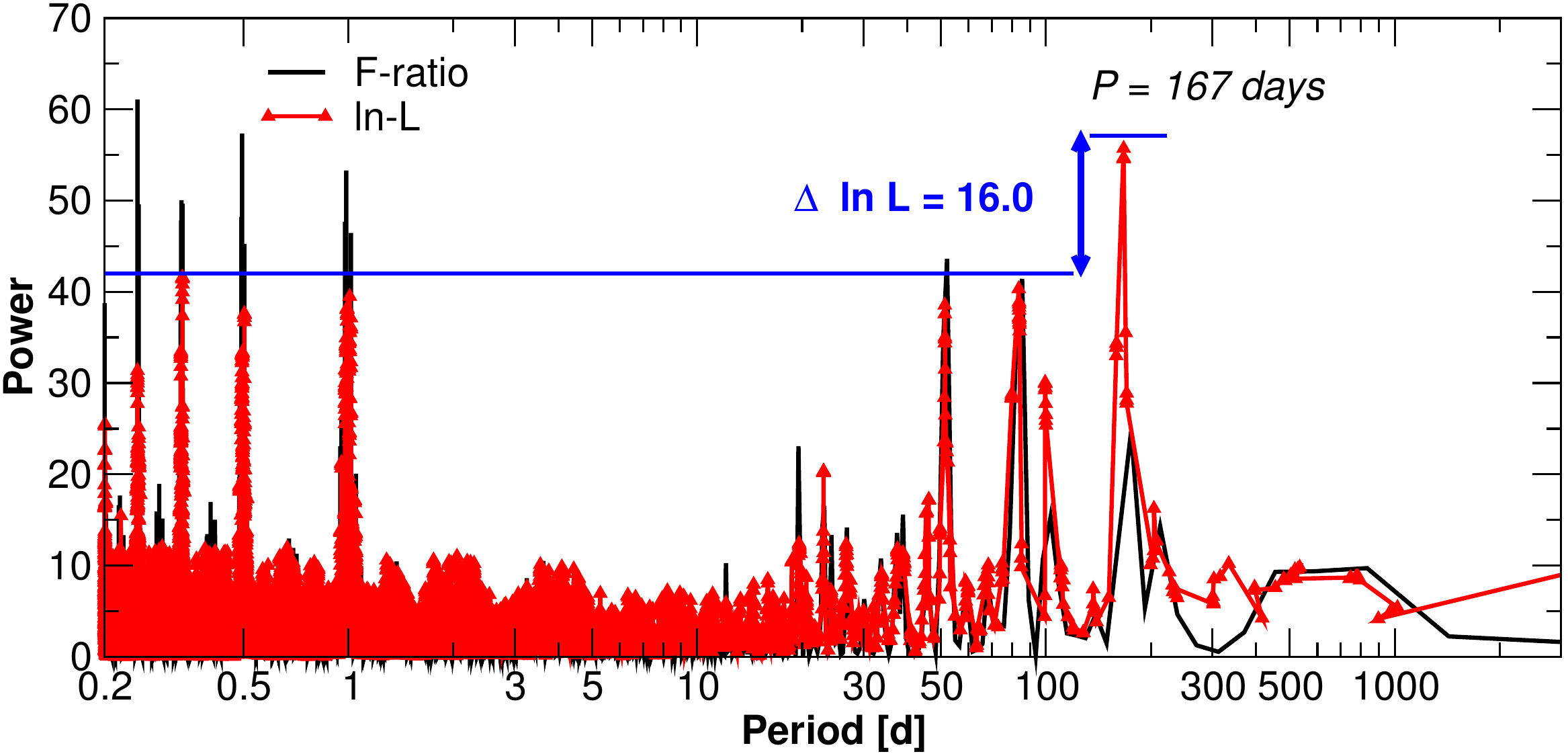}}
\caption{Likelihood periodogram of the radial velocities of the \target{}.
The black curve is a version of the generalized Lomb-Scargle periodogram for
circular orbits using the F-ratio statistic to represent the significance of each solution.
The red triangles show the improvement on the log-likelihood statistic using a full
Keplerian fit at the period search level instead.} 
\label{fig:periodogram}
\end{figure}
The analysis of the radial velocity variation is done using the generalized
Lomb-Scargle (GLS) periodogram \citep{zechmeister2009}, the likelihood
function approach of \citet{baluev2008}, and a final fit of a Keplerian orbit.
\Autoref{fig:periodogram} shows the likelihood periodogram % of the radial velocities
of the \target{} for the period range 0.2--\SI{1000}{\day}.
The black curve (F-ratio) represents the GLS periodogram for circular orbits. %, which underestimates the
% significance of eccentric solutions.
It shows highly significant periods at \Po{}, fractions of \Po{}, \Pf{},
and \Pe{}. The \Po{} period and fractions of it are aliases of our nightly observation basis.
With Keplerian fits for the same period range, the resulting picture is different. The
red triangles (ln-L curve) show these Keplerian fits as a log-likelihood
statistic. The \Ph{} period has a very low false-alarm probability of ~\num{2.2e-8},
so the signal is extremely significant. Compared to all other peaks,
the ln-likelihood of the preferred solution is higher by $\Delta \ln{L} = 16$.
Within the framework of this type of periodogram, this implies
a $\sim\num{8.9e6}$ higher probability.
Moreover, the \Pf{} and \Pe{} peaks are most likely
only harmonics of this period ($\Pe{} \approx \Ph{}/2$) and the
window function, which has a high power at \SI{135}{\day} ($1/51 -
1/83 \approx 1/135$). We also performed a detection probability test 
comparable to \citet{fiorentino2010} to verify that our sampling is
sensitive to all periods in the probed range.

\begin{figure}
\resizebox{\hsize}{!}{\includegraphics[clip, trim=0.2cm 0.1cm 0.2cm 0.2cm]{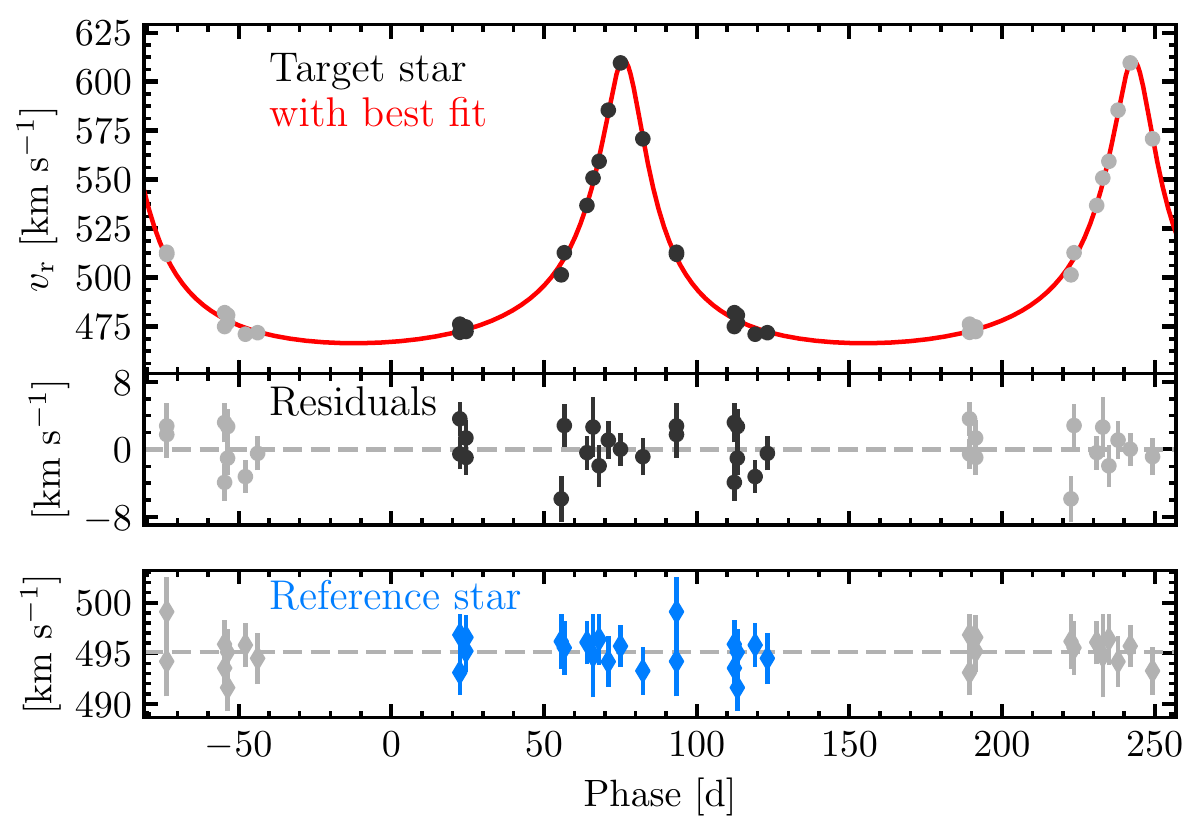}}
\caption{The top panel shows the radial velocity measurements $v_\mathrm{r}$ of the \target{}, phase folded for the \Ph{} period. Error bars are smaller than the data points. The red curve shows the best-fitting Keplerian orbit (see \autoref{table:binary_results}). The middle panel contains the residuals after subtracting this best-fitting model to the data. The bottom panel shows the radial velocity measurements of the reference star. Grey dots are phase shifted duplicates of the black ones, and are included to improve the visualization.}
\label{fig:velocity_curve}
\end{figure}
 
\begin{table}
 \caption{Binary system properties. The Keplerian parameters were calculated with an MCMC consisting of $10^7$ iterations. The derived mass was calculated using $\num{2.0e4}$ MCMC samples and the same number of \target{} mass samples.}
 \label{table:binary_results}
 \centering
 \begin{tabular}{r l}
 \hline\hline\rule{0pt}{2.3ex}
 Period $P$ & $166.88^{+0.71}_{-0.63}$ \si{\day} \\[2pt]
 Doppler semi-amplitude $K$ & \SI{69.4(25)}{\kilo\meter\per\second} \\
 Eccentricity $e$ & \num{0.595(22)} \\
 Argument of periastron $\omega$ & \SI{2.6(32)}{\degree} \\
 Periastron passage $T_0$ & \SI{57140.2(5)}{\day} \\
 Barycentric radial velocity $\gamma_0$ & \SI{494.5(24)}{\kilo\meter\per\second} \\
 Linear trend $\dot{\gamma}$ & \SI{-0.27(270)}{\kilo\meter\per\second} \\
 Jitter $s$ & $0.68^{+0.40}_{-0.25}$ \si{\kilo\meter\per\second} \\[3pt]
 \hline\rule{0pt}{2.1ex}
 Minimum companion mass $M \sin(i)$ & \BHmass{} \\
 Minimum semi-major axis $a(M)$ & \SI{1.03(3)}{\au} \\
 \hline
 \end{tabular}
\end{table}

\Autoref{fig:velocity_curve} shows the radial velocity measurements phase folded with the \Ph{} period and the best-fitting Keplerian orbit solution. The reduced $\chi^2$ of the Keplerian orbit fit is \num{1.2} (for comparison the reduced $\chi^2$ of the best circular orbit at the \Pe{} period is 45).
%(for comparison the reduced $\chi^2$ of the best circular orbit is \num{47} for the \Pe{} period). %The bottom in \autoref{fig:velocity_curve} shows the residuals of the data points to this orbital solution.
The resulting orbital parameters fitted with the MCMC approach of \citet{haario2006} are shown in \autoref{table:binary_results}. The reference star does not show any significant radial velocity variation (reduced $\chi^2 = \num{0.42}$).

%\begin{figure}
%\resizebox{\hsize}{!}{\includegraphics{plots/mass_function}}
%\caption{Mass function of the binary system. The graph shows the resulting companion mass for a given \target{} mass, and a fixed period, eccentricity and amplitude listed in \autoref{table:binary_results}. The minimum mass of the companion is even for a test mass above \SI{3}{\solarmass}.}
%\label{fig:mass_function}
%\end{figure}

\section{Discussion and conclusions}
\label{sec:discussion}
The data show strong evidence that the \target{} is in a binary system with a non-luminous object having a minimum mass of \BHmass{}. This object should be degenerate, since it is invisible and the minimum mass is significantly higher than the Chandrasekhar limit ($\sim\SI{1.4}{\solarmass}$). The small distance of the system from the centre of the \gc{} (\SI{10.8}{\arcsec}) is expected from mass segregation.
% Probably the \target{} was caught by the massive object since it has a high eccentricity. Its mass of \SI{0.81}{\solarmass} supports this hypothesis since the majority of the central cluster stars should be the most common massive stars, which are the turnoff stars in this globular cluster.
Most likely, the degenerate object has exceeded the Tolman-Oppenheimer-Volkoff limit that predicts all objects to collapse into black holes above $\sim\SI{3}{\solarmass}$ \citep{bombaci1996}. We note that the mass estimate of the dark companion depends only weakly on the mass of the \target{} within reasonable error estimates (e.g. for the unrealistic case of a \target{} with \SI{0.2}{\solarmass}, the minimum companion mass will still be above \SI{3}{\solarmass}).

Alternatively, the discovery could eventually be explained through a triple star system that consists of a compact double neutron star binary with a main-sequence turn-off star around it. In the literature, neutron star binaries show a narrow mass distribution of \SI{1.35(4)}{\solarmass} per star \citep{Thorsett1999,lattimer2012}. Recent discoveries show that a single neutron star could reach \SI{2.0}{\solarmass} \citep{ozel2016}. Thus, a double neutron star system with both components having more than \SI{2.0}{\solarmass} could explain the observations. Since such a system was not observed to date and the actual mass of the discovered object is probably higher, a black hole scenario is more likely. In this case, our results represent the first direct mass estimate of a (detached) black hole in a globular cluster.
%, hence a combination with a white dwarf with \SI{0.7}{\solarmass} could also explain the derived mass. These alternative explanations hold only for a maximum mass of \SI{4.0}{\solarmass} 
%which has due to inclination a probability of $(1 - 2 \arcsin{(3.0 / 2.7)} / \pi) \approx \SI{29}{\percent}$ for the less likely period of \SI{51.28}{\day}. All in all the probability for this alternative scenario is of the order of \SI{15}{\percent}.

The recent discovery of coalescing black hole binaries \citep{abbott2016c} suggests that there is a large population of stellar-mass binary black holes in the Universe. Our results confirm that the components of such binaries can be found in \gc{}s.

The black hole is assumed to be detached because the closest possible approach in our best-fitting model is \SI{0.4}{\au} and the Roche limit for the \target{} with a reasonable radius of $\SI{1}{\solarradius}$ is of the order of $\SI{3}{\solarradius}$. We have no evidence that the black hole accretes mass emits X-rays or radio jets.

Compared to other \gc{}s, the most unusual structural parameter of \ngc{3201} is the large cluster core radius \citep[\SI{1.3}{\arcmin}, see][]{harris}. As the presence of black holes can lead to an expansion of the core radius through interactions between black holes and stars \citep{strader2012}, the discovery of the presented black hole could be an indication that \ngc{3201} possesses an extensive black hole system in its core. More observations with MUSE could reveal more black hole companions using the radial velocity method.

%The second gravitational wave event caused by the merging of two black holes with solar masses of $14.2^{+8.3}_{-3.7}$ and $7.5^{+2.3}_{-2.3}$ found by \citet{abbott2016c} suggests that there is a large population of stellar-mass binary black holes in the Universe. We can confirm that such stellar-mass black hole components can be found in \gc{}s.

To get the true system mass function, it is necessary to measure the inclination of the system. Since the Sun's distance to the \gc{} \ngc{3201} is \SI{4.9}{\kilo\parsec} \citep{harris}, the orbital movement of the \target{} is of the order of \num{0.2}\,mas. This should be observable with interferometers. Unfortunately, for the ESO interferometer VLTI GRAVITY, a reference star with K~\SI{11}{mag} within \SI{2}{\arcsec} is missing \citep{gravity}. Maybe it could be observable with diffraction limited large telescopes like HST, the VLT with NACO at UT1, or the VLT with the new adaptive optics facility (GRAAL) at UT4. Certainly, this would be a nice astrometry task for the JWST and for the upcoming first-light instrument MICADO at the ELT \citep{micado2016}.

\section*{Acknowledgement}
BG, SD, SK, and PMW acknowledge support from the German Ministry for
Education and Science (BMBF Verbundforschung) through grants 05A14MGA,
05A17MGA, 05A14BAC, and 05A17BAA. GAE acknowledges a Gauss-Professorship
granted by the Akademie f\"ur Wissenschaften zu G\"ottingen. JB acknowledges support by
Funda\c{c}{\~a}o para a Ci{\^e}ncia e a Tecnologia (FCT) through national
funds (UID/FIS/04434/2013) and Investigador FCT contract 
IF/01654/2014/CP1215/CT0003, and by FEDER through COMPETE2020 
(POCI-01-0145-FEDER-007672). This research is supported by the German Research
Foundation (DFG) with grants DR 281/35-1 and KA 4537/2-1. Based on
observations made with ESO Telescopes at the La Silla Paranal
Observatory (see programme IDs in \autoref{table:observations}).
Based on observations made with the
NASA/ESA Hubble Space Telescope, obtained from the data archive at the
Space Telescope Science Institute. STScI is operated by the
Association of Universities for Research in Astronomy, Inc. under NASA
contract NAS 5-26555.
Supporting data for this article is available at: \url{http://musegc.uni-goettingen.de}. \\

This is a pre-copyedited, author-produced version of an article accepted for publication in Monthly Notices of the Royal Astronomical Society: Letters following peer review. The version of record, Volume 475, Issue 1, 21 March 2018, Pages L15--L19, is available online at: \url{https://academic.oup.com/mnrasl/article/475/1/L15/4810643} (DoI: 10.1093/mnrasl/slx203).

%%%%%%%%%%%%%%%%%%%%%%%%%%%%%%%%%%%%%%%%%%%%%%%%%%


\bibliographystyle{mnras}
\bibliography{bh}

%%%%%%%%%%%%%%%%%%%%%%%%%%%%%%%%%%%%%%%%%%%%%%%%%%

% Don't change these lines
\bsp	% typesetting comment
\label{lastpage}
\end{document}